\documentclass[showpacs,aps,pra,twocolumn]{revtex4}
\begin{document}

\title{Nonlocality degradation of two-mode squeezed vacuum in absorbing
optical fibers}
\author{
Radim Filip,
Ladislav Mi\v{s}ta Jr.}
\affiliation{Department of Optics,
Palack\'y University,
17. listopadu 50, 772 00 Olomouc,
Czech Republic}
\date{\today}

\begin{abstract}
Transfer of nonlocal two-mode squeezed vacuum state through
symmetrical and asymmetrical lossy channel is analysed and
we demonstrate that the nonlocality is more robust against losses,
than it has been previously suggested. It can be important
for security of continuous-variable
quantum cryptography employing entangled states.

\end{abstract}

\maketitle
\section{Introduction}

Quantum mechanical entanglement and nonlocality are the inherent
features of microworld having many practical applications in
quantum information processing and communication \cite{quantinf}.
The quantum information protocols developed originally in the
context of qubit systems have been successfully extended to
continuous-variable (CV) domain \cite{conti}.
Recently, both CV analogues of BB84 and Ekert's quantum
cryptography \cite{crypt} protocols have been suggested \cite{cvcrypt}.
In experimental implementations of CV quantum cryptography protocols with
entangled states the two-mode squeezed vacuum state
\begin{eqnarray}\label{rhonopa}
{\rho}_{\rm NOPA}&=&(1-{\lambda}_{1}^2)\sum_{m,n=0}^{\infty}
{\lambda}_{1}^{m+n}|m,m\rangle\langle n,n|,
\end{eqnarray}
where $\lambda_{1}=\tanh r$ and $\sinh^{2}(r)=\langle n\rangle_{r}$ is the mean number of signal
photons ($r$ is the squeezing parameter), commonly serves as a quantum
channel. This state (\ref{rhonopa}) can be generated either by using nondegenerate parametric
amplifier (NOPA) \cite{nopa1} or by mixing two independent single-mode
squeezed fields at a beam splitter (or in fiber coupler) \cite{teleexp,nopa2}.
The quantum key distribution goes as follows. The NOPA source emits
two fields: one is distributed to Alice (A)
and the other propagates towards Bob (B).
Alice and Bob randomly select to measure one of two conjugate
field quadrature amplitudes $Q$, $P$ and,
as the squeezing parameter $r$ increases, the correlation
(anticorelation) between the results of the same quadrature measurements
on Alice's and Bob's side grows. Communicating their choices of
the measurements through a public classical channel and keeping only the
results when both of them measured the same quadrature the key is
established between Alice and Bob.

In reality, transfer of the entanglement between two distant partners,
usually realized by means of optical fibers, suffers by undesirable losses
which strongly reduce the amount of the entanglement.
These losses can be induced either by the fiber imperfections or
they can be introduced deliberately by an eavesdropper Eve (E).
If the losses are modelled by reservoir at zero temperature, one can
show that the entanglement of the state (\ref{rhonopa}) does not vanish
completely for any extinction coefficient of fibers \cite{Scheel01}.
On the other hand,
the nonlocal correlations between communicating parties Alice
and Bob guarantee the security against the individual attacks \cite{bell_crypt}.
We employ a CV analogue of this statement and assume that quantum
channel has a sufficient quality if it preserves the nonlocality of
distributed NOPA states. Thus,
knowing the fiber link transmitivity, one can judge its convenience
before experimental implementation.

In practice, it might be based on the Bell inequalities \cite{bell,Horodecki95}
for CV systems \cite{bellconti,Wilson01,Chen}.
The nonlocality of the two-mode squeezed state
under damping is deteriorated as has been recently demonstrated
utilizing the Banaszek-W\' odkiewicz inequality \cite{Jeong00}.
Qualitatively, it was shown that for fixed fiber losses the larger the
initial squeezing is the more rapidly the nonlocality is degraded
irrespective of the entanglement preservation.
In this report, we investigate the nonlocality of the two-mode squeezed
vacuum state in vacuum environment employing more appropriate CHSH
Bell inequalities than those used in \cite{Jeong00}.
By introducing the ``spin-1/2'' operators in infinite-dimensional Hilbert
space one obtains the two-qubit form of the CV Bell inequalities
\cite{Chen} to which the Horodecki's nonlocality criterion \cite{Horodecki95}
can be directly applied. This procedure is shown to be more efficient
to test the NOPA state nonlocality even when the local losses are
involved. In the absence of losses the maximal Bell factor rapidly converges
to maximal value $B_{\rm max}=2\sqrt{2}$ with increasing squeezing
parameter $r$. If the losses are involved, then we confirm the qualitative
statement that the stronger initial squeezing induces more rapid nonlocality
degradation \cite{Jeong00}. However, it is proved that the transmitted
NOPA state does not admit a local realistic description for much more pronounced
losses than it has been pointed out in \cite{Jeong00}. This fact is important
for secure long-distance quantum communication based on the shared
two-mode squeezed vacuum state.

\section{Nonlocality degradation}

We consider a lossy two-line optical-fiber transmission channel
injected by the NOPA state (\ref{rhonopa}). Simulating the losses by
Markovian reservoir at zero temperature and eliminating free
oscillations the evolution of the NOPA state can be effectively described
by the following standard master equation for the density matrix
\begin{eqnarray}\label{master}
\frac{d\rho}{dt}&=&\Gamma_{A}(2a_{A}\rho a_{A}^{\dag}-a_{A}^{\dag}a_{A}\rho-
\rho a_{A}^{\dag}a_{A})+\nonumber\\
& &+\Gamma_{B}(2a_{B}\rho
a_{B}^{\dag}-a_{B}^{\dag}a_{B}\rho-
\rho a_{B}^{\dag}a_{B}),
\end{eqnarray}
where $\Gamma_{i}$, $i=A,B$ are damping constants and $a_{i},a^{\dag}_{i}$
are annihilation and creation operators of modes $A$ and $B$. Equivalently,
the following nonunitary evolution can induce an eavesdropper Eve
by mixing the modes $A$ and $B$ with vacua at two beam splitters with
reflectivities $R_{i}=\sqrt{1-\exp(-\Gamma_{i} t)}$, $i=1,2$.
Solving the equation (\ref{master}) for initial NOPA state (\ref{rhonopa})
one finds the following solution
\begin{eqnarray}\label{rho}
\rho&=&(1-\lambda^{2})\sum_{m,n=0}^{\infty}(\lambda\sqrt{1-R^{2}_{A}}
\sqrt{1-R^{2}_{B}})^{m+n}\times\nonumber\\
& &\times\sum_{k,l=0}^{\mbox{min}(m,n)}\sqrt{{m\choose
k}{m\choose l}{n\choose k}{n\choose l}}\times\nonumber\\
& &\times\left(\frac{R_{A}}{\sqrt{1-R_{A}^{2}}}\right)^{2k}
\left(\frac{R_{B}}{\sqrt{1-R_{B}^{2}}}\right)^{2l}\times\nonumber\\
& &\times |m-k,m-l\rangle\langle n-k,n-l|.
\end{eqnarray}
Note that if the losses are introduced only on the Bob's side ($R_{A}=0$),
the (A,E) state can be obtained from the state (\ref{rho}) by exchanges
$R_{B}\to\sqrt{1-R_{B}^{2}}$ and $\sqrt{1-R_{B}^{2}}\to-R_{B}$.

To turn the infinite dimensional nonlocality problem into the two-qubit
problem, one can simply introduce the following single-mode realization of
the Pauli matrices \cite{algebra}
\begin{eqnarray}\label{smatrix}
S_{1}&=&\sum_{m=0}^{\infty}(|2m\rangle\langle
2m+1|+|2m+1\rangle\langle 2m|),\nonumber\\
S_{2}&=&i\sum_{m=0}^{\infty}(|2m+1\rangle\langle 2m|
-|2m\rangle\langle 2m+1|),\nonumber\\
S_{3}&=&\sum_{m=0}^{\infty}(|2m\rangle\langle 2m|
-|2m+1\rangle\langle 2m+1|),
\end{eqnarray}
satisfying the Pauli matrix algebra
\begin{eqnarray}\label{commut}
[S_i,S_j]=2i\varepsilon_{ijk}
S_{k},\quad(S_i)^{2}=1,
\end{eqnarray}
where $i,j,k=1,2,3$; $\varepsilon_{ijk}$ is the totally antisymmetric tensor
with $\varepsilon_{123}=+1$. Due to the commutation rules (\ref{commut}),
the nonlocality of our state (\ref{rho}) can be investigated by means of the
standard two-qubit CHSH Bell inequalities in which the Pauli matrices are
replaced by the operators (\ref{smatrix})
\begin{eqnarray}\label{Bell}
2\ge|\langle({\bf a}\cdot{\bf S}^{A})({\bf b}\cdot{\bf
S}^{B})\rangle+\langle({\bf a}'\cdot{\bf S}^{A})({\bf b}\cdot{\bf
S}^{B})\rangle\nonumber\\
+\langle({\bf a}\cdot{\bf S}^{A})({\bf b}'\cdot
{\bf S}^{B})\rangle-\langle({\bf a}'\cdot{\bf S}^{A})({\bf
b}'\cdot{\bf
S}^{B})\rangle|,
\end{eqnarray}
where ${\bf a}$, ${\bf a}'$, ${\bf b}$, ${\bf b}'$ are unit
vectors in the real three-dimensional space and the angle brackets
denote the averaging over the matrix $\rho$. Now, according to the
Horodecki nonlocality criterion \cite{Horodecki95}, if
\begin{equation}
B_{\mbox{max}}\equiv2\sqrt{u+u'}>2,
\end{equation}
where $u$ and $u'$ are two greater eigenvalues of the matrix $U=V^{T}V$
($V_{ij}=\mbox{Tr}(\rho S_{i}^{A}S_{j} ^{B})$ and $T$ stands for the
transposition), then the density matrix (\ref{rho}) violates the
inequalities (\ref{Bell}) for some vectors ${\bf a}$, ${\bf a}'$,
${\bf b}$ and ${\bf b}'$.

Employing (\ref{rho}) and (\ref{smatrix}) one finds after some algebra,
that the matrix $U$ is diagonal with two-fold eigenvalue $\alpha^{2}$ and
single eigenvalue $\beta^{2}$
\begin{eqnarray}
\alpha&=&2(1-\lambda^{2})\sum_{m=0}^{\infty} (m+1)\times\nonumber\\
& &\times\left(\lambda\sqrt{1-R_{A}^{2}}\sqrt{1-R_{B}^{2}}\right)^{2m+1}
\Lambda_{A}(m)\Lambda_{B}(m),
\end{eqnarray}
\begin{equation}
\Lambda_{i}(m)=\sum_{k=0}^{[m/2]}{m\choose
2k}\frac{1}{\sqrt{2k+1}}\left(\frac{R_{i}}{\sqrt{1-R_{i}^{2}}}\right)
^{2m-4k},
\end{equation}
where $i=A,B$ and
\begin{equation}
\beta=\frac{1-\lambda^{2}}{1-\lambda^{2}(1-2R_{A}^{2})(1-2R_{B}^{2})}.
\end{equation}
The maximal Bell factor then reads
\begin{equation}\label{bell}
B_{\mbox{max}}=2\sqrt{\alpha^{2}+\mbox{max}(\alpha^{2},\beta^{2})}.
\end{equation}
In the following we investigate two special cases: (i) $R_{A}=R_{B}=R$
and (ii) $R_{A}=R, R_{B}=0$. The first case corresponds to the symmetric
location of Alice and Bob with respect to the source of the NOPA state,
whereas the second case is connected with asymmetric arrangement, in which
Bob resides in the close vicinity of the source.

The nonlocality analysis of the transmitted state (\ref{rho}) is based
on the numerical calculation of the value of the maximal Bell parameter
$B_{\mbox{max}}$ (\ref{bell}). We confirm the qualitative result of the previous
paper \cite{Jeong00} that the rapidity of nonlocality degradation
of the NOPA state increases with increasing initial squeezing, as can be seen
in Fig.~1. However, there is pronounced difference in the rapidity of
the nonlocality destruction in comparison with the result found in
\cite{Jeong00}. Our analysis reveals that the nonlocality is
preserved even if stronger losses are taken into account. In the
following we perform a quantitative comparison of our results with results
obtained in \cite{Jeong00}.
If $r=1$, then the Fig.~1 demonstrates that the nonlocality threshold
for the {\em symmetric} case (i) occurs for $R\approx 0.42$ in contrast to the threshold value $R\approx 0.13$
calculated in \cite{Jeong00}. In our analysis the threshold
value $R\approx 0.13$ is achieved for larger squeezing $r>2$.
Better results are obtained in the {\em asymmetric} case (ii). If $r=2$,
then the threshold is shifted to value $R\approx 0.24$ as can be
seen in Fig.~2.

The dependence of the threshold damping parameter $R_{\mbox{max}}$ on the
squeezing parameter $r$ for symmetric and asymmetric setups is depicted
in Fig.~3 and Fig.~4, respectively. The inequality $R<R_{\mbox{max}}$
represents sufficient condition for nonlocality of the NOPA state to
be preserved during the transfer through lossy optical fibers. If the
squeezing $r$ is sufficiently large $(r\geq 1.5)$, then one can find the
following fits for the {\em symmetric} case (Fig.~3)
\begin{equation}\label{app1}
R_{\rm max}\approx 1.64 e^{-r}
\end{equation}
and for the {\em asymmetric} case (Fig.~4)
\begin{equation}\label{app2}
R_{\rm max}\approx 1.2 e^{-r}.
\end{equation}
These approximative rules enable us to check simply the ability of the
lossy fiber communication channel to transfer the nonlocality of the NOPA
state.

The losses can be described by the dimensionless absorption coefficient
\begin{equation}
\gamma=\frac{\Gamma L}{v_{f}}=\ln\sqrt{\frac{P(0)}{P(L)}},
\end{equation}
where $P(0)$ is input optical power, $P(L)$ is output optical power,
$L$ is length of the fiber and $v_{f}$ is the field velocity in the
fiber. According to the results obtained in \cite{Jeong00} the initial
squeezing $r=1$ ($B_{\rm max}=2.19$) leads to the following threshold
absorption coefficient $\gamma_{\mbox{max}}\approx 8.5\times 10^{-3}$.
On the other hand, our analysis gives the following threshold absorption
coefficient $\gamma_{\mbox{max}}\approx 9.7\times 10^{-2}$. Thus, it is
important fact for the experimentalist that the nonlocality of the NOPA
state with $r=1$ is preserved at ten times longer transmission length
than it was suggested earlier \cite{Jeong00}. Thus the quantum key can
be securely distributed even under the influence of stronger
losses.
\section{Conclusion}

We analyse the nonlocality evolution of the two-mode squeezed vacuum
state transferred through symmetric or asymmetric lossy
channels. The nonlocality preservation
is important to ensure quality of quantum channel for CV
cryptography with entangled states. We confirm qualitative statement about
pronounced nonlocality degradation with increasing initial squeezing for
fixed losses. However, utilizing the stronger nonlocality criterion in the
infinite-dimensional Hilbert space, we demonstrate the nonlocality
preservation in the presence of substantially larger losses, than it was
suggested previously. Maximal allowable losses preserving the nonlocality
can be approximately expressed by the simple formulas (\ref{app1}) and
(\ref{app2}), which can be useful for the experimentalist to quickly
determine the possibility of nonlocality transfer.
The theoretical nonlocality proof presented here is more simple in
comparison with the proof based on
the generalized Banaszek-W\' odkiewicz (B-W) version of Bell
inequalities which requires the numerical maximization
in the eight-dimensional space.
On the other hand,
it is not clear as to measure the spin-like operators
(\ref{smatrix}) experimentally.
Thus, our approach is appropriate for theoretical calculations,
whereas the generalized B-W inequalities are more suitable for the
experimental test of nonlocality.

\section{Acknowledgments}
The authors would like to thank J. Fiur\' a\v sek for useful discussions.
This work was supported by project LN00A015 and project CEZ:J14/98 of the
Czech Ministry of Education and by the EU grant under QIPC, project
IST-1999-13071 (QUICOV).


\newpage
\begin{figure}
\vspace{1cm}
\vspace{1cm}
\caption{Violation of the Bell inequalities for symmetrically distributed
NOPA state in dependence on the damping parameter $R$.}
\end{figure}
\begin{figure}
\vspace{1cm}
\vspace{1cm}
\caption{Violation of the Bell inequalities for asymmetrically distributed
NOPA state in dependence on the damping parameter $R$.}
\end{figure}
\begin{figure}
\vspace{1cm}
\vspace{1cm}
\caption{Threshold losses $R_{\rm max}$ for symmetrically distributed
NOPA state in dependence on the squeezing parameter $r$: the solid line
represents exact numerical calculations and the dotted line represents the
approximative result.}
\end{figure}
\begin{figure}
\vspace{1cm}
\vspace{1cm}
\caption{Threshold losses $R_{\rm max}$ for asymmetrically distributed NOPA
state in dependence on the squeezing parameter $r$: the solid line
represents exact numerical calculations and the dotted line represents the
approximative result.}
\end{figure}

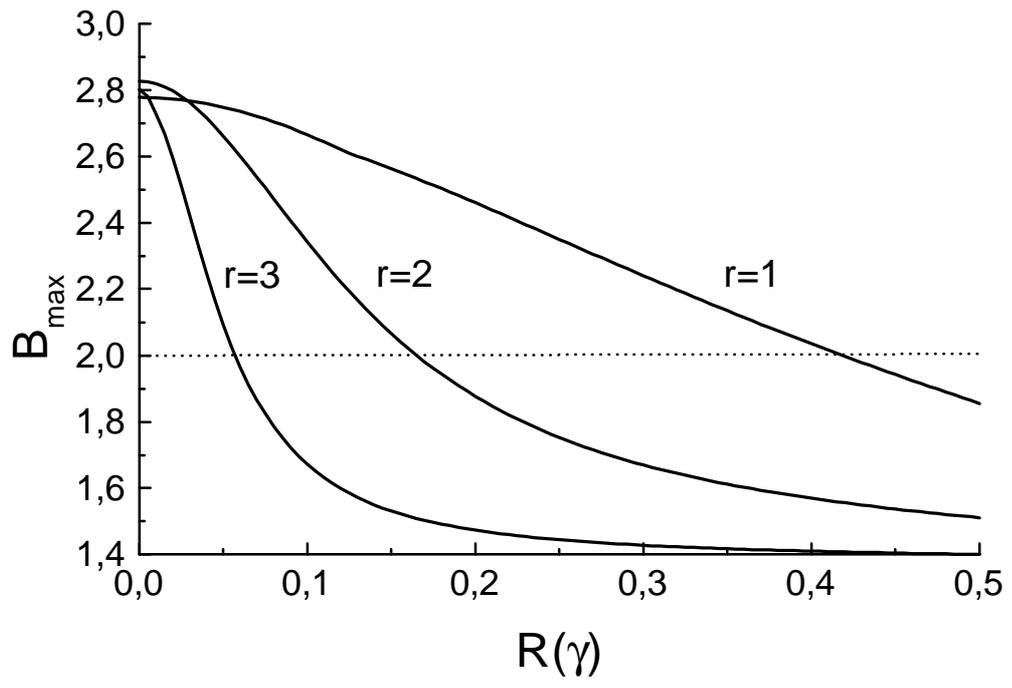
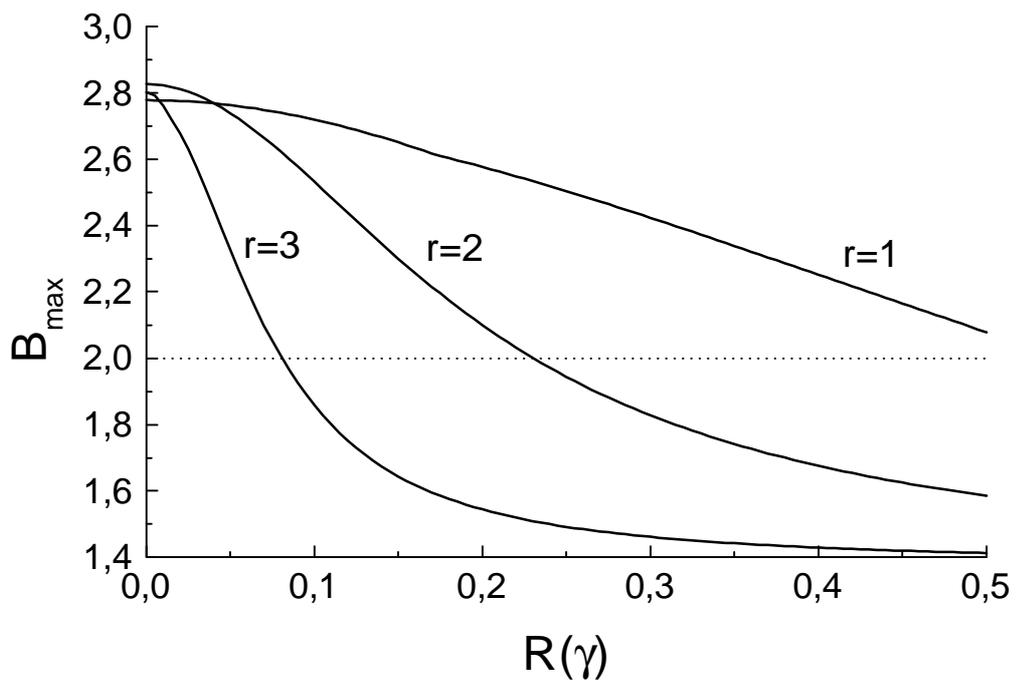
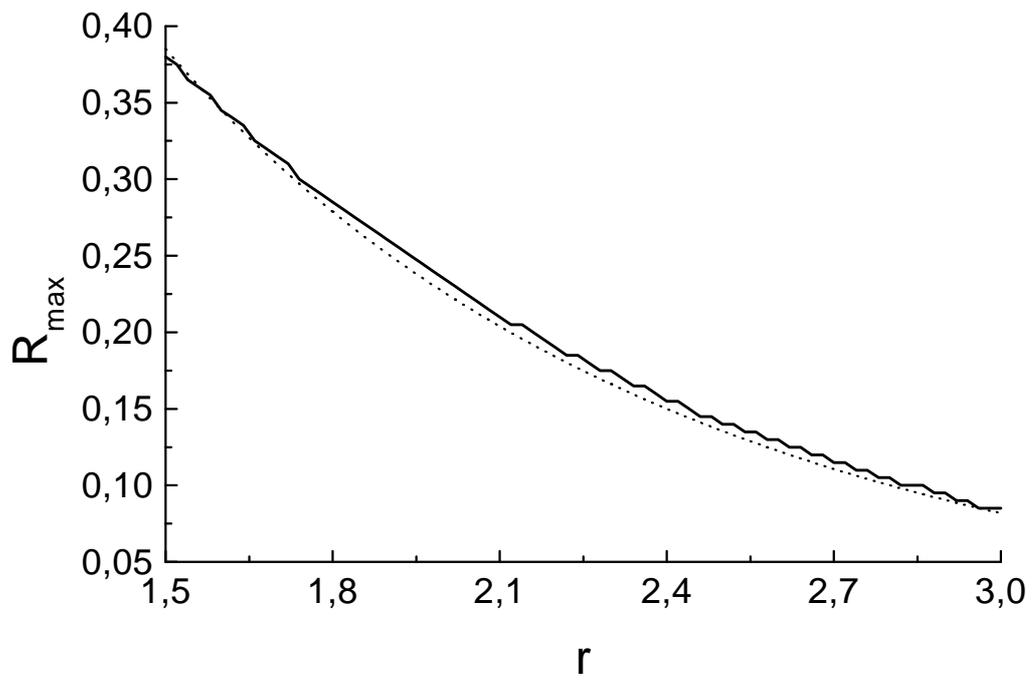
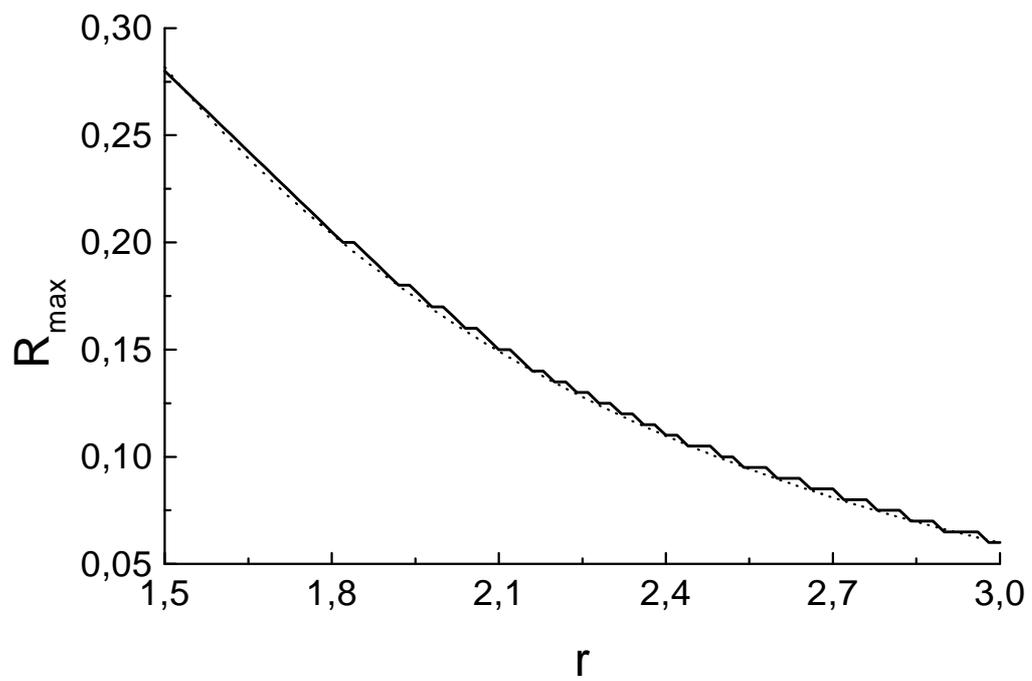
\end{document}